# Target allocation yields for massively multiplexed spectroscopic surveys with fibers

Will Saunders [a1], Scott Smedley[a], Peter Gillingham[a], Jaime E. Forero-Romero[b], Stephanie Jouvel[c], Brian Nord[d]

[a]Australian Astronomical Observatory, PO Box 915, North Ryde NSW 1670, Australia;
[b]Departamento de Física, Universidad de los Andes, Bogotá, Colombia;
[c]Institut de Ciències de l'Espai, Torre C5 par-2, Barcelona 08193, Spain;
[d]Fermilab Center for Particle Astrophysics, FNAL, Batavia, IL 60510-0500

## ABSTRACT

We present Simulated Annealing fiber-to-target allocation simulations for the proposed DESI and 4MOST massively multiplexed spectroscopic surveys. We simulate various survey strategies, for both Poisson and realistically clustered mock target samples. We simulate both Echidna and theta-phi actuator designs, including the restrictions caused by the physical actuator characteristics during repositioning.

For DESI, with theta-phi actuators, used in 5 passes over the sky for a mock ELG/LRG/QSO sample, with matched fiber and target densities, a total target allocation yield of 89.3% was achieved, but only 83.7% for the high-priority Ly-alpha QSOs. If Echidna actuators are used with the same pitch and number of passes, the yield increases to 94.4% and 97.2% respectively, representing fractional gains of 5.7% and 16% respectively. Echidna also allows a factor-of-two increase in the number of close Ly-alpha QSO pairs that can be observed.

Echidna spine tilt causes a variable loss of throughput, with average loss being the same as the loss at the rms tilt. The simulated annealing allows spine tilt minimization to be included in the optimization, at some small cost to the yield. With a natural minimization scheme, we find an rms tilt always close to $0.58 \times$ maximum. There is an additional but much smaller defocus loss, equivalent to an average defocus of 30μm. These tilt losses offset the gains in yield for Echidna, but because the survey strategy is driven by the higher priority targets, a clear survey speed advantage remains.

For 4MOST, high and low latitude sample mock catalogs were supplied by the 4MOST team, and allocations were carried out with the proposed Echidna-based positioner geometry. At high latitudes, the resulting target completeness was 85.3% for LR targets and 78.9% for HR targets. At low latitude, the target completeness was 93.9% for LR targets and 71.2% for HR targets.

**Keywords:** fiber positioners, fiber spectroscopy, simulated annealing, multi-object spectroscopy, spectroscopic surveys

## 1. INTRODUCTION

The survey speed of any massively multiplexed spectroscopic instrument depends on the fiber allocation completeness (the fraction of fibers that can be allocated to targets), and the target allocation completeness (the fraction of targets that actually get observed). Both the target and fiber allocation completeness for a multi-fiber survey instrument depend on the nature of the target samples (e.g. priorities, clustering, re-observation requirements), the physical characteristics of the fiber positioners (e.g. the patrol area for each actuator), and also the survey strategy (e.g. the number of passes over a given piece of sky). We have carried out allocation simulations for actuator geometries proposed for DESI and 4MOST, for both Poisson and more realistic mock target distributions. The allocations use Simulated Annealing (SA) [1], and this application has strong parallels with crystal formation, its original inspiration. Simulated Annealing has already been incorporated into the positioner allocation software for 2dF on the AAT [2]. It was found that as well as always finding close-to-optimal configurations, it avoids the complex selection function variations caused by more sophisticated algorithms. SA is rather slow, but computer speeds increase faster than fiber numbers, so that problem has eased and will continue to do so. The key to reasonable run time is the prior calculation and efficient indexing of all potential allocations and all pairwise allocation conflicts.

---
[1] will@aao.gov.au

This paper is concerned with the effects on the target and fiber completeness (or 'yield') of actuator geometry and target sample characteristics. The effect of the overall size and shape of the positioner was not considered, we simply assumed large square areas of actuators and targets. Unless noted otherwise, the target and total fiber densities (i.e. actuators × passes) on the sky were always matched as closely as possible, to allow clean comparisons between different geometries.

## 2. POSITIONER GEOMETRIES

The positioner was always assumed to have hexagonally spaced actuators, and sky curvature was neglected. We assumed circular patrol areas for each actuator, and a circular exclusion radius around any allocated target; this is appropriate for both $\theta$-$\phi$ and Echidna-style actuators. The actuator geometry is then defined completely by the patrol radius and the exclusion radius for each actuator, and the pitch between actuators, all in angular units on the sky.

Four basic options were considered:
1) The first represents DESI with $\theta$-$\phi$ actuators. The pitch was taken to be 2.4′ (or 10.4mm), the patrol radius 6mm, and an exclusion radius 2.25mm (all approximate values for DESI). 5 passes were assumed, which is the average DESI value. This geometry is also appropriate for the earlier BigBOSS design [3], which had a slower plate scale but very similar geometry in angular terms.
2) The second represents DESI with Echidna-type actuators, with 2.4′ (10.4mm) pitch, 10.4mm patrol radius, 0.625mm exclusion radius, and 5 passes.
3) The third represents the 'Mohawk' system proposed for DESpec [4], with a faster plate-scale, a pitch and patrol radius of 6mm, a 0.6mm exclusion radius, but only 2 passes. The overall spine×pass and target density on the sky is assumed to be identical to DESI.
4) Finally, we considered a design appropriate to 4MOST [5], with 2.4′ (9mm) pitch, 10.4mm patrol radius, 0.625mm exclusion radius. For 4MOST, there are two spine types, low resolution (LR) and high resolution (HR), with a 2:1 mix, and two corresponding target sets. 6 passes (or 'tiles' in 4MOST parlance) were assumed.

For DESI, where there are multiple passes over the sky, the home positions were dithered between passes. This is especially beneficial for $\theta$-$\phi$ positioners, with their smaller and almost discrete patrol areas.

For $\theta$-$\phi$ actuators, it was assumed that any allocation allowed by this geometry was realizable (though this is not a trivial problem in itself). For Echidna, individual moves are not precise in either length or direction, and the position is refined iteratively with metrology after each move, much like a game of golf. The requirement to avoid collisions between spines at all times leads to additional constraints, both to the individual moves and in the final allowed configurations. For the original Echidna on Subaru/FMOS [6], the patrol radius was restricted to 0.9 pitches, and any pair of allocations whose average spine tilt could be reduced by swapping targets was disallowed; in combination, these prevented collisions. However, both DESI and more especially 4MOST have a requirement for larger patrol radii, and 4MOST has two spine types (so swapping targets is often not an option), so something more sophisticated is needed.

Therefore, for each spine and for each potential target, we define 3D envelopes, within which we are sure the spine will remain, given the uncertainties in the motion. We conservatively assumed a maximum error of 25% of the requested move in any direction. We then forbid these envelopes to overlap between different spines, giving a list of pairwise exclusions of spine/target combinations. We also assume the first two moves from home position towards the target are deliberately made shorter than required, so there is no possibility of overshooting the target and colliding with other spines, as explained in Figure 1. This requires just one additional iteration over a scheme that moves straight from home to target, which would imply a much larger exclusion radius around each target.

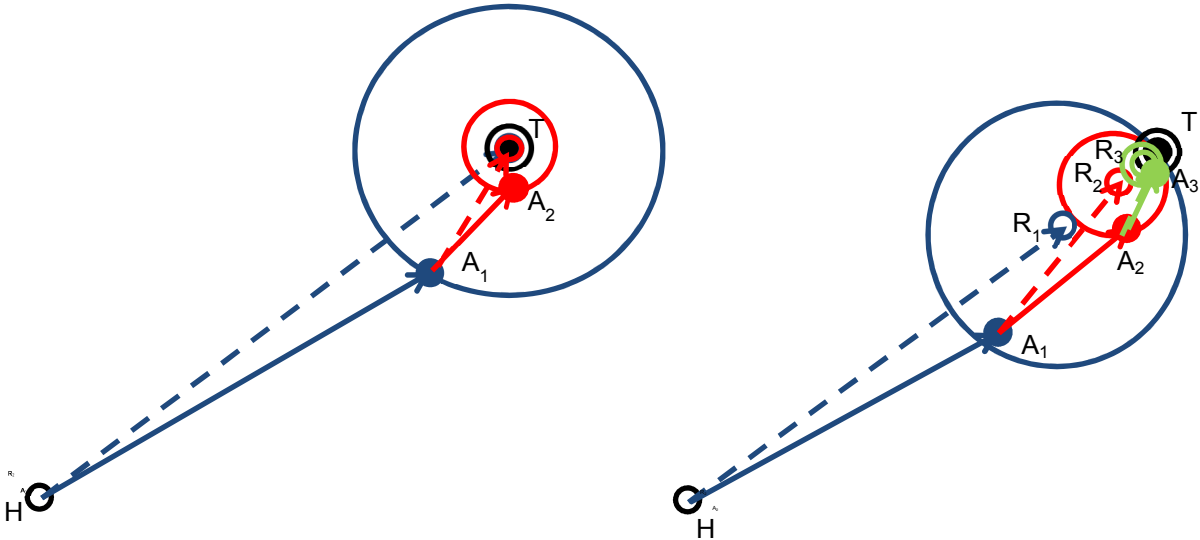

Figure 1. Schematic illustration of alternative iterative positioning algorithms. (a) On the left is the simple scheme where a spine with home position *H* is iteratively moved to a requested target position *T*. The black circle represents the physical exclusion radius (the spine tip). The large blue circle represents the maximum uncertainty in the first move (and hence the exclusion radius that would be required with this algorithm), and $A_1$ represents a (poor) first actual achieved position. A second iteration is shown in red. (b) On the right is the proposed scheme allowing close pairs to be safely configured: the requested positions ($R_1$, $R_2$, $R_3$) are always chosen such that the target lies on the edge of the uncertainty circle, and the center of that circle lies on the line *HT*. Then the spine can never overshoot, and the actual positions $A_1$, $A_2$, $A_3$ etc always remain within the first error circle, allowing close pairs of targets to be configured. The cost is that (for 25% maximum errors on a single move) an additional iteration is required to achieve the same positioning precision.

## 3. GALAXY SIMULATIONS

All galaxy simulations were matched to the positioner geometry in overall area and shape on the sky. The easiest galaxy simulations to produce are Poisson simulations. For these, the number of targets was always tuned to be the same as the total fiber availability (the product of the number of spines and the number of passes, i.e the maximum number of observations that could in principle be made), to allow consistent comparisons. Poisson simulations provide a benchmark completeness for comparison of positioner geometries, but they overestimate the completeness for real samples, because clustering, differing priorities, and re-observation requirements all reduce the yield.

For more realistic simulations for DESI and DESpec, we have used the mock galaxy catalogs based on the algorithm "Adding Density Determined GAlaxies to Lightcone Simulations" (ADDGALS, [7], [8]). This algorithm attaches synthetic galaxies, including multiband photometry, to dark matter particles in a light-cone output from a dark matter N-body simulation and is designed to match the luminosities, colors, and clustering properties of galaxies. The catalog used here was based on a single `Carmen' simulation run as part of the LasDamas of simulations [9]. This simulation modelled a flat ΛCDM universe with $\Omega_m = 0.25$ and $\sigma_8 = 0.8$ in a 1 Gpc/$h$ box with $1120^3$ particles. A 220 sq deg light cone extending out to z = 1.33 was created by pasting together 40 snapshot outputs. Color cuts, similar to those used in BOSS and SDSS, were made to generate ELG and LRG samples respectively, but at the depth and target mix expected for DESI; the full prescription is in the appendix. The number of targets was tuned to be approximately the same as the total fiber availability, to allow consistent comparisons. For the purpose of the comparison between positioner geometries, we simply took a 3° × 3° square piece of this, containing 21800 ELGs and 3249 LRGs (the latter each requiring 2 observations). QSO candidates were added at random (Poisson-distributed), again at the density expected for DESI, and were split into 2 types;1440 potential Ly-α absorption targets requiring observation on every pass, and 900 others requiring only a single observation. For the Ly-α targets, 37.5% were presumed confirmed after the first observation, with the rest no longer requiring re-observation (being stars or QSOs with $z<2.2$), and the complete allocation was redone for the remaining 4 passes. These numbers were all taken to match the expectations laid out in the DESI CDR. The total number of desired target observations (including re-observations) was 32798, for the 5 passes with 6525 spines. The target distributions are shown in Figure 2.

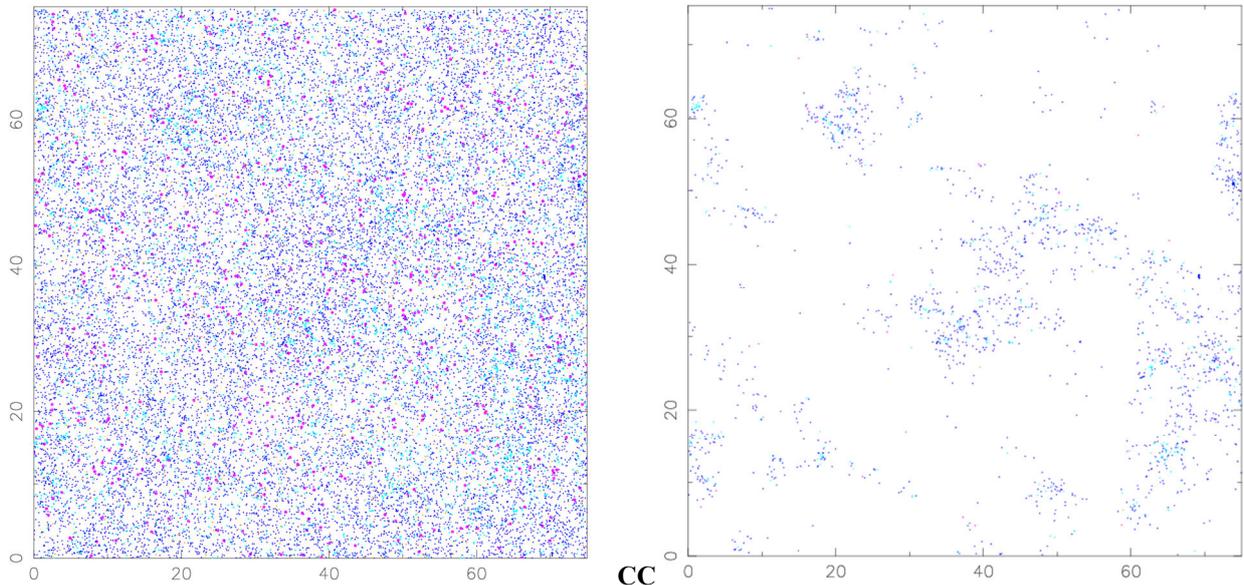

Figure 2(a) Sky distribution of the sample used for testing DESI allocations. ELGs are blue, LRGs are light blue. Tracer QSOs are orange, good Ly-α targets are magenta, and bad Ly-α targets are yellow. Symbol size is scaled by the number of observations required. The box is 3°×3°, and the axes are in units of the pitch between actuators (10.4mm or 2.4′). 2(b) unallocated targets after 5 passes.

For 4MOST, simulated 5°×5° test catalogues for both high and low latitude were supplied by the 4MOST team and were used as given. In both cases, we assumed that 6 passes (or tiles) would be made, with 20 minutes integrations, giving a total of 58240 potential LR observations and 29120 potential HR observations. The target counts are always much less than the total spine availability, partly because 6 tiles is not expected to be always achieved, and partly because there will be other large community science target lists to be added in. The high latitude sample contains 6 LR and 2 HR target types, with widely varying priorities and re-observation requirements. One of the LR target sets is very highly clustered (galaxies in x-ray selected clusters). There are 21551 LR targets, requiring 35289 observations (60.6% of the available spine × tile count). There are 2189 HR targets, requiring 13937 observations (only 47.9% of the available spine × tile count, but all targets requiring multiple re-observation). The target distribution is shown in Figure 5.

At low latitude for 4MOST, the situation is simpler. There are only 2 target types - 48633 LR targets each requiring a single observation (83.5% of the available spine × tile count), and 2567 HR targets requiring 15669 observations (53.8% of the available spine × tile count). The target distribution is shown in Figure 6.

## 4. ALLOCATION SIMULATIONS

The allocation algorithm used Simulated Annealing (SA) to find the optimum allocation. This has already been implemented in the allocation algorithm for 2dF [2], and the tiling algorithm for 6dFGRS [10] A nice feature of SA is that the merit function can be tuned in any way wanted; so for example different target types can be given different priorities, or large actuator radii can be penalized, or the number of sky fibers optimized.

In our implementation, all fibers start off de-assigned. Assignments and de-assignments are then made at random, with assignments always made to the best available target, and with the probability of de-assignment depending on both target merit and time[2]. This is done by assigning a 'temperature' $T$ to the system, and accepting de-assignments with probability $exp$ $(-1/T)/m_i$, where $m_i$ is the merit of the $i$th target to be observed[3]. $1/T$ was linearly increased, from $1/(1000sp)$ to $1/10$, where $s$ is the number of spines and $p$ is the number of passes. $10^5 sp$ trial assignments or de-assignments were made in total. Trial with $10^6 sp$ re-assignments showed only a very small increase in the completeness.

The key to doing these huge numbers of reassignments in reasonable computation times lies in the efficient tabulation and indexing of all the possible actuator-target assignations, and of all the exclusions that prevent any given reassignment. These exclusions may be because (a) the proposed target is already allocated, (b) that it is too close to another allocated target in the same pass, or (c) that this actuator-target assignment would conflict with an existing actuator-target assignment in the same pass. These exclusions include huge ($\sim 10^{18}$ element) but sparse ($10^6$ non-zero element) matrices, so indexing is mandatory. The calculation of these matrices currently marginally dominates the run-time, but it needs doing only once for a given catalogue and positioner geometry, and can be done off-line. The multiple sets of passes are annealed simultaneously, so pairwise exclusions can always be circumvented by having targets allocated on different passes, and the annealing does appear to reliably do this.

For DESI, only a minority of the Ly-α QSO candidates turn out to actually be so, and so the entire field is reconfigured accordingly after each pass. But until a Ly-α QSO candidate is observed, there is still a choice whether to assume it will be good (and so will require re-observation, or to assume it will be bad (and so not require re-observation). We have here assumed the former, which gives better yields for the Ly-α QSOs, at some cost to the other targets. It affects all geometries in the same way, so does not strongly affect the relative results.

Where sources require re-observation, some reflection of this is needed in the weighting. Our solution is, for any proposed allocation, to decrease the weights by a factor $0.75^{(n_{req}-n_{obs})}$, were $n_{req}$ is the number of observations required for that source, and $n_{obs}$ is the number it would have if this allocation is made. This ensures that completed sets of allocations are more advantageous than multiple partially-completed sets.

For both surveys, it is expected that some fields will fail to get the desired number of passes, for various reasons. Therefore allocations on earlier passes are more valuable than those on later passes. This was reflected by decreasing the weights by a factor $0.9^p$, where $p$ is the pass (1st, 2nd etc) for the allocation in question.

## 5. RESULTS

The results for the various tests for DESI and DESpec are given in Table 1, while Figure 3 shows a small sample of an allocation for DESI+Echidna, showing the exclusion 'cones' for each spine/target pair, and Figure 2(b) shows the unallocated targets.

Table 1. Completeness results for DESI or DESpec for various geometries, strategies and target samples. Results in boldface represent our best estimate of the yields of Echidna and $\theta$-$\phi$-based positioners in actual use for DESI.

| Actuator geometry | Target catalog | Method | # passes | Total Yield | ELG | LRG | Ly-α QSO | Tracer QSO |
|---|---|---|---|---|---|---|---|---|
| DESI + $\theta$-$\phi$ | Poisson | SA | 5 | 95.19 | | | | |
| DESI + Echidna | Poisson | SA | 5 | 97.95 | | | | |
| DESpec + Mohawk | Poisson | SA | 2 | 94.78 | | | | |
| **DESI + $\theta$-$\phi$** | **Mock** | **SA** | **5** | **89.31** | **89.10** | **89.12** | **83.74** | **95.44** |
| **DESI + Echidna** | **Mock** | **SA** | **5** | **94.42** | **92.97** | **94.74** | **97.22** | **97.89** |
| DESpec + Mohawk | Mock | SA | 2 | 91.44 | 91.75 | 87.44 | 97.86 | 97.99 |

---

[2] Direct reassignments from one target to another were not permitted, because of the enormous increase in book-keeping it entails. However, direct reassignment would greatly decrease the required number of trials (and hence the run-time), and would surely be included in any actual implementation of the code into an observatory system.

[3] A probability of $exp(-m_i/T)$ was used initially in the code, but this has the undesirable feature of almost completely decoupling the allocations for higher and lower weight targets, for convenient weights like 1,2,3.

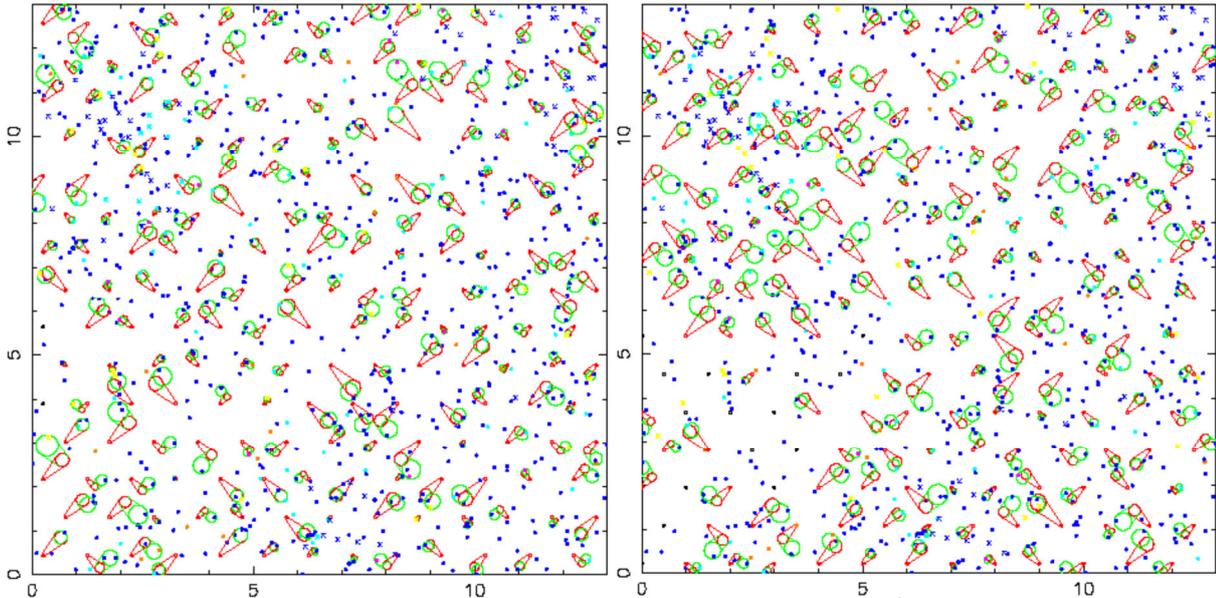

Figure 3. Sample DESI+Echidna simulation, 5 passes, 25% errors, (a) 1st pass, (b) 2nd pass. Color scheme for the targets is as for Figure 2. Filled circles represent targets with sufficient allocations overall, while crosses have insufficient observations. Symbol size shows the number of observations if completed, or the number of missing observations if not (so big filled circles are good, big crosses are bad). Open black circles show unused spines. The red and green cones represent the uncertainties in the coarse positioning movements of the spine on the first and second portions of the move, and are not allowed to overlap with other cones of the same color. The 1st pass has a much higher fraction of Ly-α targets, both good (magenta) and bad (yellow).

For DESI, Echidna always gives better completeness than $\theta$-$\phi$, especially for the higher priority targets (those requiring first-pass or multiple observations). The overall improvement in yield is 5.7% (94.4% vs 89.3%), with a gain of 16% (97.2% vs 83.7%) for the high priority Ly-α QSOs. There is a much larger gain, about a factor of 2, in the number of close pairs (~1′ separation) of good Ly-α targets, with both observed. These close pairs are invaluable, in that they allow cross-sightlines of similar lengths to the scales being tested for the effects of the neutrino mass; this then allows a test for systematic error in the results from the much more common line-of-sight Ly-α absorber pairs; this systematic error is expected to dominate the error budget for this determination. The re-observation requirement for the Ly-α QSOs has a very large impact on the target completeness, for both $\theta$-$\phi$ and Echidna geometries. In general, *any* constraints or modifications to the merit function, can only decrease the overall yield.

The unallocated targets are overwhelmingly associated with large-scale surface density variations in the galaxy distribution (Figure 2b). This means that (a) effect of the incompleteness on the selection function can be readily modelled by mapping these density variations, and (b) the 3D density of observed targets is more uniform than the surface density.

With a DESpec+Mohawk -type system with only 2 sky passes instead of 5, the completeness is still 91.4% overall, and 97.9% for the Ly-α QSOs. However, the Ly-α QSOs only receive 2 observations instead of 5.

We tested the yield for plausible variations to the Echidna geometry or performance – 3-axis instead of 2-axis actuators, and 11% or 43% maximum errors on the moves, instead of 25%. These large changes made remarkably small changes to the yield (typically 1-2%), though the biggest gains were always for the highest-value targets.

Overall fiber yields for realistic samples were always less than 95%, leaving a satisfactory number for sky fibers[4].

---

[4] The formal optimum number of sky fibers, assuming $N$ fibers per spectrograph, $M$ degrees of freedom in the sky spectrum, and no systematic errors, is $\sqrt{NM}$. Different sky-subtraction methods assume $M$ from 1 (e.g. BOSS's

The results for 4MOST at high-latitude are shown in Figure 4 and quantified in Table 2. For most samples, excellent completeness (80%+) was achieved. The exceptions were the very highly clustered X-ray-selected sample, where a completeness of 46.6% was achieved[5] – still enough to give many redshifts per cluster, allowing dynamical mass estimation; and also the HR targets, almost all of which require re-observation on every tile, and so have low completeness, just like the DESI Ly-α QSOs.

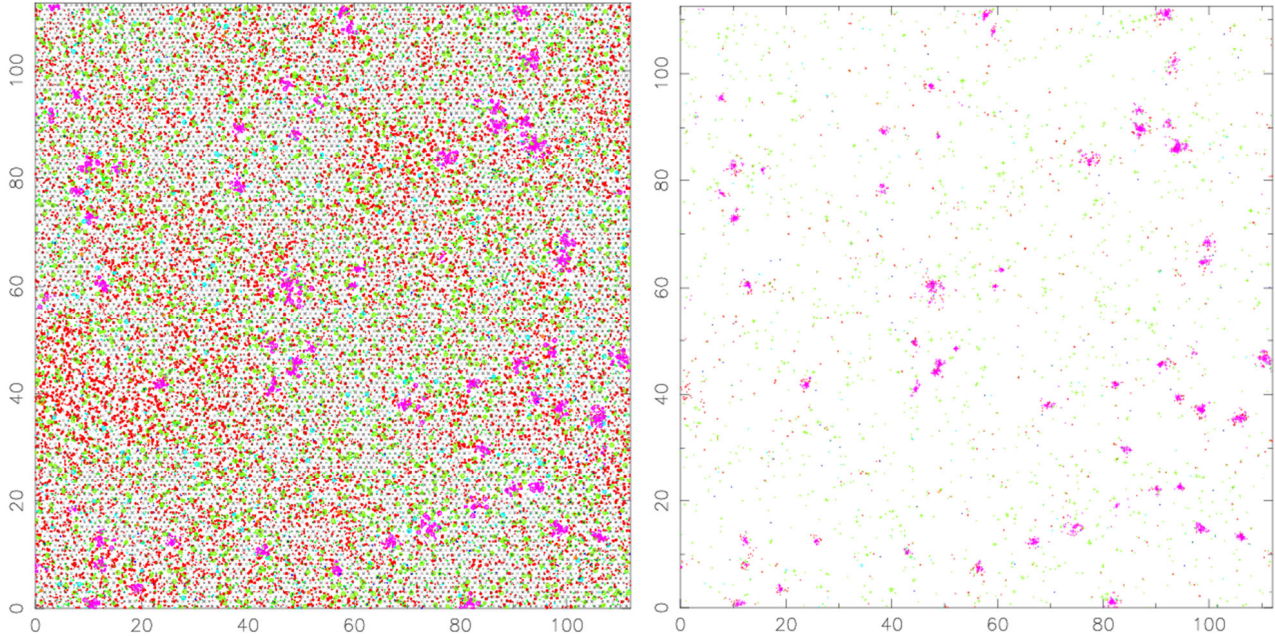

Figure 4. (a) allocated and (b) unallocated (or not fully allocated) 4MOST targets at high latitude in 6 passes (or tiles). Units are actuator pitches (9.4mm or 2.4′) . Symbol size is (a) the number of observations or (b) the number of allocations short of the desired number. HR targets are colored pale green and orange. The magenta sources are galaxies in rich galaxy clusters, with an allocation completeness of 46.6%. Overall completeness (actual observations vs desired) is 85.3% for LR targets and 52.0% for LR spines, and 78.9% for HR targets and 37.8% for HR spines.

The results for 4MOST at low latitude are shown in Figure 5 and Table 2. There is significant incompleteness caused by the large target density variation, and again reduced completeness for the HR targets because of the re-observation requirements. The overall LR completeness, at 93.9%, is close to that demanded in any case by sky fiber requirements.

Table 2. Completeness results for 4MOST simulations.

| Sample | LR fiber yield | LR target yield | HR fiber yield | HR target yield | AGN | BAO | Clusters | Halo LR | Disk LR | Halo HR | Disk HR |
|---|---|---|---|---|---|---|---|---|---|---|---|
| Poisson, matched target numbers | 96.6 | 96.6 | 71.3 | 71.3 | | | | | | | |
| High latitude DRS mock | 52.0 | 85.3 | 37.8 | 78.9 | 94.5 | 94.8 | 46.6 | 97.3 | 82.0 | 86.7 | 78.7 |
| Low latitude DRS mock | 77.6 | 92.9 | 38.3 | 71.2 | | | | | 92.9 | | 71.2 |

---

'spectroperfectionism', [11]) to a few (the Principal Component Analysis sky subtraction used e.g. for AAOmega). The integrated S/N has a rather flat maximum, and 5% seems a reasonable minimum for ~500 fiber spectrographs.

[5] It may be possible to improve this number further, using a variant of the manouvering shown in Figure 1, that allows close triplets of targets (as opposed to just pairs) to be simultaneously and safely observed.

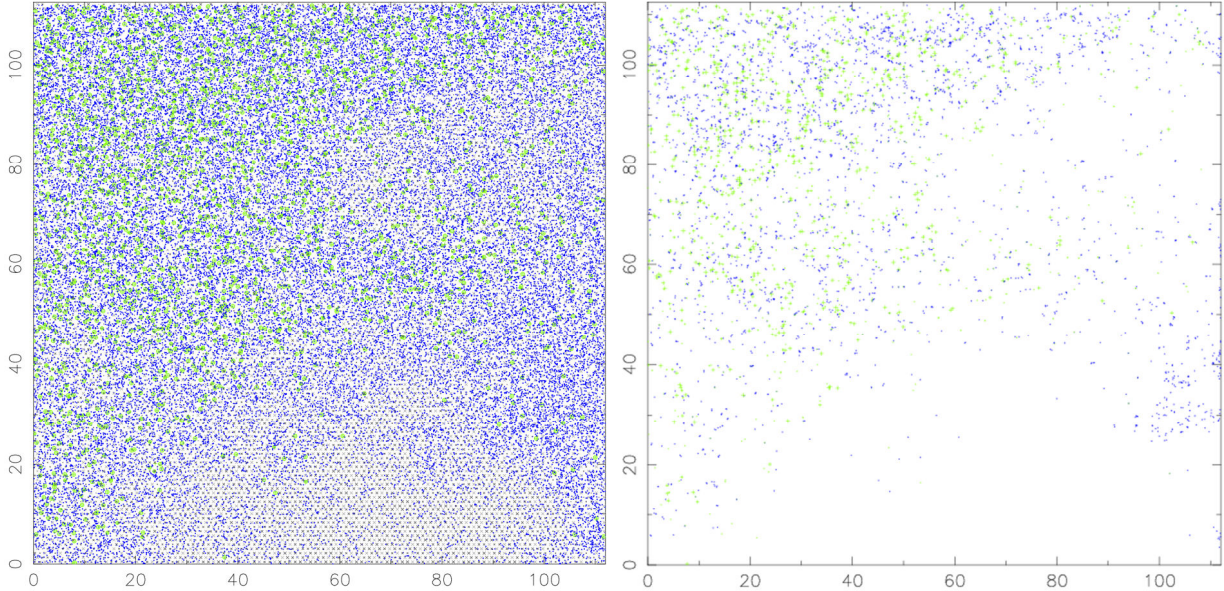

Figure 5. (a) allocated and (b) unallocated (or not fully allocated) 4MOST targets at low latitude in 6 passes (or tiles). Units are actuator pitches (9.4mm or 2.4′). Symbol size is (a) the number of observations or (b) the number of allocations short of the desired number. LR targets are blue, HR targets are pale green. Overall completeness (actual observations vs desired) is 93.9% for LR targets and 77.6% for LR spines, and 71.2% for HR targets and 38.3% for HR spines.

## 6. TILT DISTRIBUTION

For Echidna-style actuators, there is a position-dependent throughput loss due to tilt-induced Focal Ratio Degradation. Also, correct focus can only be set for one tilt, and any other tilt causes an additional aperture loss due to defocus. Both these losses have a quadratic dependence of tilt, which means that both the average tilt losses and the best focus value are given by the rms tilt (rather than the simple average). Because of this, tilt minimization for DESI was performed assuming a quadratic penalty function, with the weights multiplied by $(1-t^2)$, where $t$ is tilt in units of 0 to 1 (at maximum tilt). This weighting greatly reduces the average tilt, at a cost of just ~ 0.7% in the completeness, to give a significant gain in overall survey speed. Figure 6(a) shows the resulting radius distributions for the mock DESI catalog, for ELGs, LRGs and QSOs.

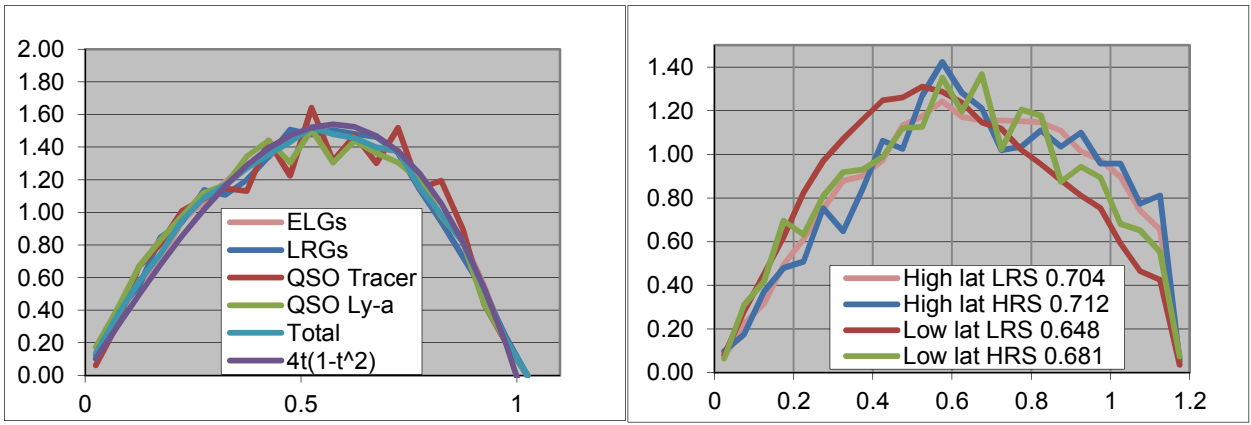

Figure 6 (a). Spine tilt distribution for a DESI allocation simulation, together with a simple cubic that well describes the data. The units on the x-axis are pitches between actuators. The rms tilt is 0.577 pitches or 1.38°.

Figure 6 (b). Spine tilt distribution for 4MOST allocation simulations, for LR and HR spines, at both high and low latitudes. The units are again the pitch between actuators. Rms tilt is shown in each case.

The tilt distribution is reasonably well fitted by the simple and physically motivated approximation $n(t)=4t(1-t^2)$, with no obvious difference between the samples. This formula gives an average tilt of 0.533, and rms (and also median) tilt of 0.577. The rms variation in focus is 0.137 × the maximum change, or 30μm. So the average throughput (averaged over all fibers) is the same as for a fiber at a tilt of 0.577 pitches (1.38°), with a defocus of 30μm.

These weights will affect the target completeness, as a function of target position and local target density. However, this can be modelled, or just measured empirically; and is much simpler than the weights already needed for other sources of variation in the completeness (spine tilt, radius on the plate, seeing, moon phase etc).

For 4MOST, a weaker weighting $(1-t^2)/2$ was used, because the two spine types mean large tilts cannot always be avoided. Because of this, the resulting tilt distribution is peaked to higher values. Figure 6(b) shows the tilt distribution for both HR and LR spines, for both high and low latitude samples. The rms tilt (in units of the pitch) is also shown. The average tilt in degrees is ~1.4°, depending somewhat on the sample, but with a broader distribution than for DESI.

## 7. DISCUSSION

Spine tilt leads to throughput losses because of (a) additional collimator losses arising from geometric Focal Ratio Degradation, and (b) aperture losses due to defocus (in general a much smaller effect). The tilt values presented here allow these losses to be quantified [12], with the result that the tilt losses amount to an average loss of ~3.2% in the S/N per target for fixed integration time (or ~6.5% loss of survey speed to a fixed S/N), with small dependencies between 4MOST and DESI, and also between target types within each survey.

For DESI, the tilt losses must be set against the allocation benefits presented above, in comparing the overall survey efficiency of Echidna and $\theta$-$\phi$ positioners. Such a comparison requires a complete survey simulation, and is outside the scope of this paper. But we note that (a) the DESI survey strategy is driven by the highest-priority targets, and these also have the largest allocation yield benefits; and (b) for the ELGs, the integration times are longer than in general needed, so the loss of S/N per target is usually irrelevant, while the allocation benefit remains. So Echidna appears to give a clear, if unquantified, overall benefit to the survey efficiency.

For 4MOST, the target allocation yields are mostly excellent. The exceptions are (a) the HR targets were multiple reobservation is always required, this combined with the lack of overlaps between HR spine patrol areas inevitably means most close pairs of targets cannot be observed (analogously to the DESI Ly-α QSOs with $\theta$-$\phi$ actuators); and (b) the cluster galaxies, where practical configuring considerations reduce the target yield below the simple expectation based on patrol and exclusion radii. However, the latter may be amenable to significant improvement with a more sophisticated configuring algorithm. The latest Echidna test results [12] suggest that that should be feasible, within the 4MOST reconfiguring time budget.

## 8. APPENDIX: SELECTION OF THE DESI ELG AND LRG SAMPLES

The Carmen simulation gave a dark-matter distribution to be populated by galaxies. The galaxy distribution was created by first using an input luminosity function to generate a list of galaxies, and then adding the galaxies to the dark matter simulation using an empirically measured relationship between a galaxy's magnitude, redshift, and local dark matter density, $P(\delta_{dm}|M_r,z)$ - the probability that a galaxy with magnitude $M_r$ and redshift $z$ resides in a region with local density $\delta_{dm}$. This relation was tuned using a high resolution simulation combined with the SubHalo Abundance Matching technique that has been shown to reproduce the observed galaxy 2-point function to high accuracy [12,13,14]. For the galaxy assignment algorithm, we choose a luminosity function that is similar to the SDSS luminosity function [15], but evolves in such a way as to reproduce the higher redshift observations (e.g., SDSS-Stripe 82, AGES, GAMA, NDWFS and DEEP2). In particular, $\phi_*$ and $M_*$ are varied as a function of redshift in accordance with the recent results from GAMA [16]. Once the galaxy positions have been assigned, photometric properties are added. Here, we use a training set of spectroscopic galaxies taken from SDSS DR5. For each galaxy, in both the training set and simulation, we measure, the distance to the 5th nearest galaxy on the sky in a redshift bin, $\Delta_5$. Each simulated galaxy is then assigned an SED based on drawing a random training-set galaxy with the appropriate magnitude and local density, $k$-correcting to the appropriate redshift, and projecting onto the desired filters. When doing the color assignment, the likelihood of assigning a red or a blue galaxy is smoothly varied as a function of redshift in order to simultaneously reproduce the observed red fraction at low and high redshifts as observed in SDSS and DEEP2.

To select the LRGs, we do a color-color cut based on the SDSS selection [17], but using DES photometry. Two color cuts were used, to select samples tuned to $0.2 < z < 0.4$ and $0.4 < z < 0.6$. To select the ELGs, we follow Escoffier et al. [18] with two sets of *gri* color cuts to define brighter ($19 < i < 21.3$, $z \sim 0.8$) and fainter ($21.3 < i < 23$, $z \sim 1.0$) samples, while avoiding both the stellar sequence and the quasar sequence.

## ACKNOWLEDGEMENTS

We are extremely grateful to Risa Wechsler and Michael Busha for permission to use their simulations to create the DESI/DESpec mock catalogs, and to Tom Dwelly, Olivier Schnurr and the whole 4MOST team for providing test sub-samples of the compiled mock catalogs for the 4MOST Design Reference Surveys.